\documentclass[11pt]{article}

\usepackage{graphicx}
\usepackage{bbm}

\newcommand{\be}{\begin{equation}}
\newcommand{\ee}{\end{equation}}
\newcommand{\bqa}{\begin{eqnarray}}
\newcommand{\eea}{\end{eqnarray}}
\newcommand{\bqas}{\begin{eqnarray*}}
\newcommand{\eeas}{\end{eqnarray*}}

\newcommand{\non}{\nonumber}

\def\brc{\langle}
\def\ckt{\rangle}

\def\D{\mathcal{D}}
\def\Tr{\hbox{\rm Tr}}

\def\Im{\hbox{\rm Im}}

\def\Tr{ \hbox{\rm Tr}}

\begin{document}


\thispagestyle{empty}

\begin{flushright}
 IFUP-TH/2008-23, \\
{\tt hep-th/yymmnnn} \\
September  2008 \\
\end{flushright}
\vspace{3mm}

\begin{center}
{\Large \bf
Advent of Non-Abelian Vortices and Monopoles\footnote{Contribution to  "30 Years of Mathematical Methods in High Energy Physics -- In honor of Professor Tohru Eguchi's 60th Birthday --", 
(Kyoto) March 2008. 
To appear as the Conference Proceedings in Progress of Theoretical Physics Supplement. 
} 
 \\
-- further thoughts about duality and confinement 
   }
\\[12mm]
\vspace{5mm}

\normalsize
  { \bf
Kenichi Konishi}
\footnote{e-mail: konishi(at)df.unipi.it}

\vspace{3mm} 

{\it Department of Physics, ``E. Fermi'',  University of Pisa \\
and 
\\
 INFN, Sezione di Pisa, \\
Largo Pontecorvo, 3, Ed. C, 56127 Pisa, Italy
} \\
%
%
%
\vspace{5mm}

{\bf Abstract}  \\[5mm]

{\parbox{13cm}{\hspace{5mm}
Recent development on non-Abeliann vortices and monopoles is reviewed with an emphasis on their relevance on confinement and duality.  A very recent construction of non-Abelian vortices which do not dynamically Abelianize is crucial in this context. 

}}
\end{center}

\section{Introduction}

It has become customary to think of confinement in QCD as a kind of dual superconductor, in which charged objects such as the quarks are confined by a
chromo-electric vortex, in a medium in which magnetic monopoles are condensed. 
But then there is no convincing evidence that the effective magnetic system is an Abelian theory. We must consider seriously the possibility that the vacuum of QCD is a dual superconductor of non-Abelian type.  One must understand the behavior of  monopoles and their interactions and vortices (confining strings), all of non-Abelian variety. The tool we shall rely upon in studying this matter will be: hints from exactly solvable systems with N=2, 1, ... supersymmetries,  study of semi-classical soliton vortices and monopoles, considerations based on the topological and symmetry argument, and the effective actions describing the dynamics of these vortices.  The aim of my talk is twofold. In the first part I shall review the state of the art of the non-Abelian vortices and monopoles, leading to the very recent developments, emphasizing the relation between the monopole and vortex moduli,  and  their importance in the physics of duality and confinement in QCD.  In the second part, I shall discuss a very recent work done in collaboration with K. Ohashi and D. Dorigoni on non-Abelian vortices which do not Abelianize dynamically.  This result is fundamental in this context. 

\section{Difficulty of earlier ideas on non-Abelian monopoles} 

Soon after the discovery of the 't Hooft-Polyakov monopoles \cite{TH}  in a spontaneously broken 
$SU(2) \to U(1)$ theory, the construction was generalized to the cases in which the symmetry breaking is of the type
\be
 G   \,\,\,{\stackrel {v_{1}} {\longrightarrow}} \,\,\, H  \,\,
\ee 
where $H$ is some non-Abelian gauge group \cite{NAmonop,GNO,BS,EW}. 
It was found that the regular monopoles arising in the above system could be characterized by the charges $\beta$  such that 
\begin{equation}   F_{ij} =  \epsilon_{ijk} B_k = 
\epsilon_{ijk}  \frac{ r_k 
}{      r^3}  ({ \beta} \cdot  {\bf H}),        \end{equation}
in an appropriate gauge, where ${\bf H}$ are the diagonal generators of $H$ in the Cartan subalgebra. 
A straightforward generalization of the Dirac's quantization condition leads to 
\begin{equation}  2 \, {\beta \cdot \alpha} \in  { \bf Z}   \label{naqcond}
\end{equation}
where $\alpha$ are the root vectors of $H$.
The constant vectors $\beta$  (with the number of components equal to the rank of the group $H$)  label possible monopoles. 
It is easy to see that the solution of Eq. (\ref{naqcond})  is  that  $\beta$  is any  of the {\it weight vectors} of a group whose  
nonzero roots are given by 
\be   \alpha^{*} = \frac{\alpha}{\alpha \cdot \alpha}. 
\label{dualg}\ee

 The group generated by Eq. (\ref{dualg}) is known as the {\it dual}   (we shall call it GNOW dual below) of  $H$,  let us call ${\tilde H}$.  One is thus led to  a set of semi-classical {\it degenerate}  monopoles, with multiplicity   equal to that of a representation of ${\tilde H}$;  this has led to the so-called  GNOW   conjecture, {\i.e.}, that they form a multiplet of the group ${\tilde H}$,   dual of  $H$ \cite{GNO}-\cite{EW}.   
 For simply-laced groups (with the same length of all nonzero roots) such as $SU(N)$, $SO(2N)$, the dual of $H$ is basically the same group, 
except that the allowed representations tell us that 
\be     U(N) \leftrightarrow  U(N);  \qquad   SO(2N)\leftrightarrow  SO(2N) ,  \ee
while 
\be  SU(N) \leftrightarrow  \frac{SU(N)}{{\mathbf Z}_{N}}; \qquad   SO(2N+1)   \leftrightarrow  USp(2N).   \label{below}
\ee

  There are however well-known difficulties with such an interpretation.  The first concerns the topological obstruction discussed
in \cite{CDyons,DFHK,Houghton,Baisbis}:  in the presence of the classical monopole background, it is not  possible  to define a globally  well-defined set of generators isomorphic to $H$.  As a consequence, no  ``colored dyons''  exist.  In a simplest case with 
the breaking 
\be  SU(3)  \,\,\,{\stackrel {\brc \phi_{1} \ckt    \ne 0} {\longrightarrow}} \,\,\, SU(2) \times U(1),
\label{simplebr}\ee
   this means  that 
    \be { no\,\, monopoles\,\, with \,\, charges }  \quad   ({\underline 2},  1^{*})  \quad  { exist},    \label{cannot} \ee
   where the asterisk indicates a dual, magnetic charge.  

The second  can be regarded as an infinitesimal version of the same difficulty:   certain bosonic zero modes around the monopole solution, corresponding to $H$ gauge transformations,  are non-normalizable (behaving as $r^{-1/2}$ asymptotically).  Thus the standard procedure of quantization leading to  $H$ multiplets of monopoles  does not work.    Some progress on the check of GNOW duality along this orthodox line of thought has  been reported  nevertheless \cite{DFHK},  in the context of  ${\cal N}=4$  supersymmetric gauge theories.  Their approach,  however, requires the consideration of particular class  of  multi monopole systems,  neutral with respect to the non-Abelian  group  (more precisely, non-Abelian part of)  $H$ only. 

Both of these difficulties concern  the transformation properties of the  monopoles  under the subgroup  $H$, while the   relevant question should be  how they transform under the dual group, ${\tilde H}$.  As field transformation groups, $H$ and ${\tilde H}$  are relatively nonlocal, the latter  should look like a nonlocal transformation group  in the original, electric description.

\section{Light non-Abelian monopoles}

In spite of these apparent difficulties, light non-Abelian monopoles do appear in the low-energy effective action of a wide class of ${\cal N}=2$ gauge theories with matter hypermultiplets.  $SU(N)$, $SO(N)$, $USp(2N)$  theories with quark multiplets have been analyzed in detail, and their occurrence as the massless low-energy degrees  of  freedom carrying non-Abelian 
dual gauge charges  have been established \cite{APS,HO,CKM}.  Some of the most salient features are
the following: 
\begin{itemize}
  \item  Renormalization-group effects on the behavior of the  monopoles are clearly understood.  For instance,  in $SU(N)$ theory, the low-energy, magnetic 
  \[  SU(r) \times U(1)^{N-r}
  \]
 gauge group  appears for  
 \be r \le  N_{f}/2  \label{range}\ee 
  only \cite{APS,HO,CKM}.   $N_{f}$ pairs of  monopoles  in the fundamental representation of $SU(r)$  help attenuate the dual gauge interactions.   The infrared degrees of freedom in such a vacuum is given in Table\ref{tabnonb}. 
  
 \item  Monopoles acquire flavor quantum numbers  in the fundamental representation of $SU(N_{f})$  via the Jackiw-Rebbi mechanism: the loop effects due to such light monopoles explain  the range of the values of $r$.  The sign flip in the dual gauge group beta function as compared to the underlying theory  is fundamental in understanding the appearance of non-Abelian monopoles as light degrees of freedom \cite{Konishi04}. 
 
 \item  Indeed, only Abelian monopoles appear in pure  ${\cal N}=2$, that is $N_{f}=0$, theories \cite{curves, curvesbis},  or in the case of $SU(2)$ gauge theories  \cite{SW1,SW2}. In the latter,  the requirement of asymptotic freedom (strong interactions in the infrared) leads to the condition $N_{f}\le 3$.  Therefore  $r\le 1$ and non-Abelian monopoles cannot appear in the infrared, and indeed do not.  
  
\end{itemize}

One of the most important lesson one learns from these theories is the fact that non-Abelian dual gauge groups (and associated monopoles) occur only in models with flavors.  This is a first hint that the dual gauge group is intimately related to the presence of a flavor symmetry.  The role of massless flavor on the appearance of non-Abelian duality is 
actually even subtler \cite{Duality}. 

The vacua $r= N_{f}/2$ (in the  $SU(N)$ supersymmetric QCD) constitute an interesting, limiting class of theories: they are infrared fixed-point theories (SCFT)
\cite{AD,AGK}.

As non-Abelian monopoles exist quantum mechanically as the low-energy effective degrees of freedom,  there must be ways to {\it understand} them somehow, in spite of the difficulties mentioned before. 

Below we shall study the monopoles   in terms of {\it  vortices}, by putting the low-energy $H$  gauge system in the Higgs phase \cite{ABEK,Duality}.   A systematic study of non-Abelian vortices started only recently,  but they seem to be much better understood than the non-Abelian monopoles.

\begin{table}[h]
\begin{center}
\vskip .3cm
\begin{tabular}{ccccccc}

&   $SU(r)  $     &     $U(1)_0$    &      $ U(1)_1$
&     $\ldots $      &   $U(1)_{n-1}$    &  $ U(1)_B  $  \\
\hline
$n_f \times  q$     &    ${\underline {\bf r}} $    &     $1$
&     $0$
&      $\ldots$      &     $0$             &    $0$      \\ \hline
$e_1$                 & ${\underline {\bf 1} } $       &    0
&
1      & \ldots             &  $0$                   &  $0$  \\ \hline
$\vdots $  &    $\vdots   $         &   $\vdots   $        &    $\vdots   $
&             $\ddots $     &     $\vdots   $        &     $\vdots   $
\\ \hline
$e_{n-1} $    &  ${\underline {\bf 1}} $    & 0                     & 0
&      $ \ldots  $            & 1                 &  0 \\ \hline
\end{tabular}
\caption{\footnotesize The effective low-energy degrees of freedom and their quantum numbers at the confining vacuum characterized by a magnetic dual $SU(r)$ gauge group.   }
\label{tabnonb}
\end{center}
\end{table}

\section{Understanding monopoles through vortices }  

The monopoles and vortices are  closely related to each other,  through the homotopy map and by a symmetry consideration. 
The moduli and non-Abelian transformation properties among the monopoles follow from those of the low-energy vortices which confine them.  When the full theory 
with hierarchical symmetry breaking 
\be
 G   \,\,\,{\stackrel {v_{1}} {\longrightarrow}} \,\,\, H  \,\,
 \,{\stackrel {v_{2}} {\longrightarrow}} \,\,\,
 {\bf 1}, 
  \label{hierarchy}
\ee 
is considered, the vortex orientational modes, which fluctuate and propagate along the vortex length and in time, get turned into dual gauge fluctuations of the monopoles  at the end, unless they become strongly coupled and dynamically Abelianize. 
The central observation is that the properties of the monopoles induced by the breaking 
\be   G \to H 
\ee
are closely related to  the properties of  the vortices, which develop when the low-energy $H$ gauge theory is put in Higgs phase by a set of scalar VEVs,
$H \to  {\bf 1}$, through 
the  exact homotopy sequence, 
\be    \cdots \to   \pi_{2}(G)   \to    \pi_{2}(G/H)   \to      \pi_{1}(H)  \to   \pi_{1} (G) \to \cdots    \label{homotopy}
\ee
Clearly regular monopoles (non-trivial elements of  $\pi_{2}(G/H)$)  and vortices of the low-energy theories (whose winding is quantized by  $\pi_{1}(H)$)    are related, because  for any compact Lie groups $\pi_{2}(G) ={\bf 1}$.   The precise relation is as follows.  If   $\pi_{2}(G) $ and  $\pi_{1} (G) $ are both trivial,    then  {\it each}   element of  $\pi_{1} (H) $  is an image of a corresponding element of   $\pi_{2} (G/H)$:   all monopoles are regular, 't Hooft-Polyakov monopoles.  No vortices  of the low energy theory is stable and they end at the regular monopoles.  Vice versa, no regular monopoles  exist in the full theory: their magnetic flux is carried away by a thin vortex line.   

\begin{figure}
\begin{center}
\includegraphics[width=3in]{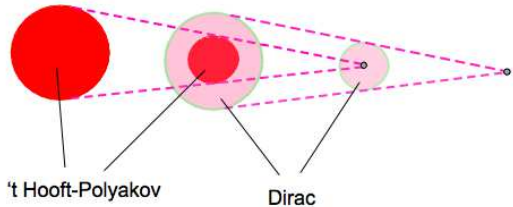}
\caption{\small A pictorial representation of the exact homotopy  
sequence,  (\ref{homotopy}),  with  the  leftmost figure  
corresponding to  $\pi_{2}(G/H)$.} \label{sequence}
\end{center}
\end{figure}

Consider instead the case $\pi_{1} (G) $ is nontrivial. Let us take for concreteness $G=SO(3)$,  with $\pi_{1}(SO(3))={\mathbf Z}_{2}$, and 
$H= U(1)$, with $\pi_{1}(U(1))={\mathbf Z}.$    The exact sequence illustrated in  Fig. \ref{sequence} in this case implies that the monopoles, classified by $\pi_{1}(U(1))={\mathbf Z}$  can further be divided into 
two classes, one belonging to the image of $\pi_{2}(SO(3)/U(1))$ -- 't Hooft-Polyakov monopoles! -- and  those which are not related to the breaking -- but to the singular, Dirac monopoles, which could be introduced in the underlying $G$ theory.  The correspondence is two-to-one: the monopoles of magnetic charges  $2\, n$ times ($n=1,2,\ldots$)  the Dirac unit are regular monopoles while those with charges $2\, n +1$ are Dirac monopoles. 
In other words,  the regular monopoles correspond to the  kernel of the map $  \pi_{1}(H)  \to   \pi_{1} (G) $  (Coleman \cite{Coleman}). 

The vortex of the low-energy $H$ theory develops a non-Abelian flux moduli, when some exact symmetry (typically a color-flavor diagonal symmetry, see below) $H_{C+F}$ is broken by individual vortex configurations.  The associated Nambu-Goldstone modes propagate inside the vortex and fluctuating in time,  turn into the  dual gauge degrees of freedom of the 
the regular monopoles at the end of the vortex.  The monopoles are now endowed with the corresponding moduli. 

\section{Concrete example:  softly broken ${\cal N}= 2$ $SU(N+1)$ theory}

To test our ideas against a concrete model, consider ${\cal N}=2$  supersymmetric gauge theories  with  various ``quark'' multiplets.  These models have a great advantage that 
by appropriately tuning the bare masses of the matter fields and adjoint field,  it allows us to study both the fully quantum mechanical light monopole systems (by using the Seiberg-Witten curves \cite{SW1}-\cite{curvesbis}  and their singularity structures) and semiclassical vortices and monopoles (by the standard semi-classical constructions).  These two different regimes are related by the holomorphic dependence of physics on the mass parameters typical of supersymmetric theories. Therefore these two pictures must match. 

  The $N=1$ chiral and gauge superfields $\Phi= \phi \, + \, \sqrt2 \,
\theta\,\psi + \, \ldots \, $, and $W_{\alpha} = -i \lambda \, + \, \frac{i
}{ 2} \, (\sigma^{\mu} \, {\bar \sigma}^{\nu})_{\alpha}^{\beta} \, F_{\mu \nu} \,
\theta_{\beta} + \, \ldots $ are both in the adjoint representation of
the gauge group, while the hypermultiplets are taken in the
fundamental representation of the gauge group.  The Lagrangian takes the form, 
 \be
L=     \frac{1}{ 8 \pi} \Im \, \tau_{cl} \left[\int d^4 \theta\,
\Phi^{\dagger} e^V \Phi +\int d^2 \theta\,\frac{1}{ 2} W W\right]
+ L^{(quarks)} + \Delta L,    \label{lagrangianGeneral}
\ee
\[ L^{(quarks)}= \sum_i \, [ \int d^4 \theta\, \{ Q_i^{\dagger} e^V
Q_i + {\tilde Q_i}^{\dagger}  e^{ {\tilde V}}    {\tilde Q}_i \} +
\int d^2 \theta
\, \{ \sqrt{2} {\tilde Q}_i \Phi Q^i    +      m_i   {\tilde Q}_i    Q^i   \}
\label{lagquark}
\]
describes the $n_{f}$ flavors of hypermultiplets (``quarks''),  
$
\tau_{cl} \equiv  \frac{\theta_0 }{ \pi} + \frac{8 \pi i }{ g_0^2}
$   is the bare $\theta$ parameter and coupling constant.
The $N=1$ chiral and gauge superfields $\Phi= \phi \, + \, \sqrt2 \,
\theta\,\psi + \, \ldots \, $, and $W_{\alpha} = -i \lambda \, + \, \frac{i
}{ 2} \, (\sigma^{\mu} \, {\bar \sigma}^{\nu})_{\alpha}^{\beta} \, F_{\mu \nu} \,\theta_{\beta} + \, \ldots $ are both in the adjoint representation of
the gauge group, while the hypermultiplets are taken in the
fundamental representation of the gauge group.

We    consider      small {\it
generic}      nonvanishing
   bare masses   $m_{i}$   for the hypermultiplets
(``quarks''), which is consistent with ${\cal N}=2$ supersymmetry.    
Furthermore   it is convenient to introduce the mass for the adjoint scalar multiplet 
\be
\Delta   L=   \int \, d^2 \theta \,\mu  \,\Tr \, \Phi^2
\label{N1pert}
\ee
which breaks supersymmetry to ${\cal N}=1$.   
    An advantage of doing so  is that  all flat directions   are eliminated  and one is left with a finite number of isolated vacua;  keeping track
of this number  (and the symmetry breaking pattern in each of them)  allows us to make  highly nontrivial check of our
analyses and find out the fate of the semi-classical vortices and monopoles in the region of strong coupling.

When the low energy $H$ gauge system is put in Higgs phase, by a smaller VEV, 
 the dual  system  is in confinement  phase.  
 The transformation  law of the  monopoles follows from that of  monopole-vortex mixed configurations in the system with a large hierarchy of energy scales, $v_{1}\gg  v_{2}$, Eq.~(\ref{hierarchy}), 
under an unbroken, exact color-flavor diagonal symmetry $H_{C+F}$.  This last symmetry is broken by individual soliton vortex,  so the latter develops continuous moduli.

In our concrete model (${\cal N}=2$ SQCD with  $N_{f}$ quark hypermultiplets)   the
underlying   
 gauge symmetry can be taken {\it e.g.} to be  $SU(N+1)$,   which is broken at a much larger mass scale ($v_{1}   \sim |m_{i}|$)     as
\be    SU(N+1)  \,\,\,{\stackrel  {v_{1}    \ne 0} {\longrightarrow}} \,\,\, \frac{ SU(N) \times U(1)}{{\mathbf Z}_{N}}.
\ee
The unbroken gauge symmetry is completely broken at a lower mass scale, 
$v_{2}\sim |\sqrt{\mu  m}|$,      as in Eq. (\ref{squarkvev}) below.

In the construction of the approximate monopole and vortex  solutions  we can consider only the VEVs and fluctuations around them which satisfy
\be   [\Phi^{\dagger},   \Phi]=0, \qquad    Q_i   =  {\tilde Q}^{\dagger}_i,  \label{trunc1}
\ee
In order to keep the hierarchy of the gauge symmetry breaking scales,   we
choose the masses such that
\[     m_{1}=\ldots = m_{N_{f}}= m, \label{equalmass}  \]
\be    m   \gg  \mu    \gg   \Lambda.  \label{doublescale}\ee
We choose to study the vacuum where
(see \cite{CKM} for the detail)
\be    \brc\Phi  \ckt =  - \frac{1}{\sqrt 2}  \, \left(\begin{array}{cccc} m & 0 & 0 & 0 \\0 & \ddots & \vdots  & \vdots \\0 & \ldots   & m & 0 \\0 & \ldots & 0 & - N \, m \end{array}\right);   \label{adjointvev}
\ee
\[
Q=  {\tilde Q}^{\dagger}  = \left(\begin{array}{ccccc}  d & 0 & 0 &  0 & \ldots \\0 & \ddots  & 0 & \vdots & \ldots \\0 & 0 & d &  0 & \ldots \\0 & \ldots & 0 &  0 & \ldots\end{array}\right),  \qquad   d = \sqrt{ (N+1)\, \mu \, m}. \label{squarkvev} \]
This is a particular case of the so-called (classical)  $r$ vacuum, with $r=N$.  
This vacuum  is in a color-flavor locked phase with 
$SU(N)_{C+F}$    global symmetry.

If one neglects the smaller squark VEV, Eq. (\ref{squarkvev})  
 the symmetry breaking  Eq. (\ref{adjointvev})  gives rise to regular magnetic monopoles with mass of order of $O(\frac{v_{1}}{g})$,  as  $\pi_{2}(G/H) \sim \pi_{1}(H) = \pi_{1}(U(1)) ={\mathbf Z}$.   The continuous transformation property among these monopoles  is our main concern.

\section {Vortex in the $U(N)$ model with $N_{f}=N$}

At scales much lower than  $v_{1} = m$ but still neglecting the smaller squark  VEV
$v_{2} =   d = \sqrt{ (N+1)\, \mu \, m} \ll  v_{1}$,    the theory reduces to an $SU(N)\times U(1)$  gauge theory \cite{ABEKY}  with $N_{f}$ light quarks $q_{i}, {\tilde q}^{i}$ (the first $N$ components of the original quark multiplets $Q_{i}, {\tilde Q}^{i}$).    In the most studied case,   $N_{f}=N$, the light squark fields can be expressed as $N\times N$  color-flavor  mixed matrix.

 The adjoint scalars are fixed to its VEV,  Eq. (\ref{adjointvev}), with small fluctuations around it,
\[ \Phi =  \brc\Phi  \ckt  (1 +    \brc\Phi  \ckt^{-1} \, {\tilde  \Phi} )  , \qquad   |{\tilde  \Phi}| \ll m.
\label{small}\]
In the consideration of the vortices of the low-energy theory,  they will be in fact replaced by the constant VEV.  The presence of the small terms Eq. (\ref{small}), however, makes the low-energy vortices not strictly BPS  (and this will be  important in the consideration of their stability below).

The quark fields are replaced,  consistently with Eq. (\ref{trunc1}),  as
\[    {\tilde q} \equiv   q^{\dagger}, \qquad   q \to  \frac{1}{\sqrt{2}} \, q,
\]
where the second replacement brings back the kinetic term to the standard form.

We further replace  the singlet coupling constant and the $U(1)$  gauge field 
appropriately: 
the net effect is
\be  {\cal L} =  \frac{ 1}{ 4 g_N^2}  (F_{\mu \nu}^a)^2  + \frac{ 1}{ 4 e^2}  ({\tilde F}_{\mu \nu})^2  +
 \left|{\cal D}_{\mu}
q \right|^2
-    \frac{e^2}{2} \, |
 \, q^{\dagger} \,  q   -     c \, {\mathbf 1}  \, |^2 - \frac{1}{2} \, g_N^2 \,| \,
 \,  q^{\dagger} \, t^a q \,  |^2,
\ee
\be  c=   N(N+1)  \sqrt {    2\, {  \mu \, m}    }  .
\ee
Neglecting the small  terms left implicit, this  is  an  $U(N)$  model,  studied widely   except for the fact that $e \ne g_{N} $ here\cite{HT,ABEK,HT2,Eto:2006pg,Tong,SY,GSY,Duality}..  

 The transformation property of the vortices can be determined  from the moduli matrix\cite{seven,Duality}.
Indeed, the system possesses BPS saturated vortices described by the linearized equations
\[
\left(\D_1+i\D_2\right) \, q = 0,
\]
\[
F_{12}^{(0)} + \frac{e^2}{2} \left( c \,{\bf 1}_N - q\, q^\dagger \right) =0; \qquad F_{12}^{(a)} + \frac{g_{N}^2}{2}\, q_{i}^\dagger  \, t^{a}\, q_{i}  =0.
\]
The matter equation can be solved exactly as in
\cite{Isozumi:2004vg,Etou,Eto:2006pg}  ($z = x^1+ix^2$) by setting
\[
q  = S^{-1}(z,\bar z) \, H_0(z),\quad
A_1 + i\,A_2 = - 2\,i\,S^{-1}(z,\bar z) \, \bar\partial_z S(z,\bar z),
\]
where $S$ is an  $N \times N$ invertible matrix  over whole of the $z$ plane, and  $H_{0}$ is  the  moduli matrix, holomorphic in $z$.

Each single vortex solution  breaks the color-flavor symmetry as
\be   SU(N)_{C+F} \to  SU(N-1) \times U(1),
\ee
leading  to the moduli space of the minimum vortices  which is
\be 
 {\cal M} \simeq  {\bf C}P^{N-1} = \frac{SU(N) }{ SU(N-1) \times U(1)}.
\ee
The fact that this moduli coincides with the moduli of the quantum states of an  $N$-state quantum mechanical system,  is a first hint that the monopoles appearing at the endpoint of a vortex, transform as a fundamental multiplet ${\underline N}$ of a group $SU(N)$.   Actually the vortex represented by the moduli matrix    (we consider here the vortices of minimal winding,  $k=1$)
\be   H_{0}(z)   \simeq  \left(\begin{array}{cccc}   1 & 0 & 0 & -a_1 \\   0 & \ddots & 0 & \vdots \\  0 & 0 & 1 & - a_{N-1} \\   0 & \ldots  & 0 & z\end{array}\right),
\label{minSUN}\ee
 can be shown explicitly \cite{Duality} to transform as a fundamental multiplet of $SU(N)$.

\subsection{ Note on the non-BPS character of the monopoles and voritces}

Our discussion based on the concrete supersymmetric models exploits the fact that the vortices (in the low-energy approximation) 
and monopoles (in the high-energy approximation) are both BPS saturated,  hence stable, in the respective effective theory. Actually, both types of solitons  cease to be BPS,  when the small corrections arising from the symmetry breaking  $G \to H$  is taken into account.\cite{Duality,AEW}. This fact, that these solitons are almost BPS but not exactly so, is of course crucial in the homotopy-map argument:  it is their meta-stability which allow them to be related to each other. 
At the same time,   the attributes characterized by integers such as the transformation property of certain configurations being multiplet of a  non-Abelian group (an exact symmetry of the full system)  cannot receive renormalization.  This is similar to the current algebra relations of Gell-Mann which are not renormalized.  CVC of Feynman and Gell-Mann also hinges upon an analogous situation.   The absence of ``colored dyons''  \cite{CDyons} mentioned earlier  can also  be interpreted in this manner.    The results obtained in the BPS limit  (in  the  limit  $v_{2}/v_{1} \to 0$)  are  thus valid at any  finite values of $v_{2}/v_{1}$ \cite{Duality}.

\section{Recent developments}

The properties of the non-Abelian vortices \cite{HT,ABEKY}  in the  $U(N)\sim SU(N)\times U(1)$ system has been studied extensively\cite{ABEK,HT2,Eto:2006pg,Tong,SY,GSY,Duality}.  The last couple of years have seen more advanced analyses on these solitons, the analysis being developed  in various different directions:
 
\begin{itemize}
  \item The detailed study of moduli and transformation   property of higher winding 
vortices\cite{seven,Duality} has been performed.   These papers contain basic results on the vortex transformation properties which allow them to be interpreted in a simple group-theoretical language.  Of course, such a property is central in understanding the dual gauge symmetry along the line indicated above. 

  \item Another development concerned the systems with larger number of matter flavors \cite{HashiTong,SYSemi,SemiL}.  As in the $U(1)$ Higgs systems with more than one ``electron'' fields,     they develop new types of moduli, the so-called semi-local  vortices.
    As the latter is related to  the 
  vacuum moduli itself, which has much richer  structure in the non-Abelian cases,  the corresponding vortex moduli are much more interesting here.   In particular, a new types of (Seiberg-like) duality has been  unearthed,   between different models having related vortex moduli  (and with the common  sigma model limit) \cite{SemiL}.
  
  \item   A systematic study of non-BPS vortices has been initiated \cite{AEW}.  Unlike the Abelian non-BPS vortices, their interactions depend on the relative orientations.  and an important result about the true nature of vortices in the systems with hierarchical symmetry breaking  Eq.~(\ref{hierarchy}), as they are  necessarily non-BPS, as emphasized in the preceding section.
  
  \item   A result of particular significance is the extension of the analysis to systems with generic gauge groups \cite{KF,GFK,AGG}. 
   As compared to the $U(N)$  models  studied in most papers, the systems based on e.g.,  $SO(N)$, $USp(2N)$ (and other) 
groups are characterized by  larger vacuum degeneracy than the $U(N)$ systems, even after the  ``color-flavor locked'' vacuum is chosen to study the solitons. This reflects the fact the complexified $U(N)$ group is just the most general linear group $GL(N, C)$ while the complexification of other groups leads to some subgroup of  $GL(N, C)$.   
Various interesting consequences of these facts, such as the fractional vortices, are being worked out at present. 
 
\end{itemize}

%


\section{Vortices which do not dynamically Abelianize}

    The vortices found in the $U(N)$ theory with $N_{f}=N$ flavors are non-Abelian in the sense that 
they carry continuous non-Abelian orientational moduli. They are massless  Nambu-Goldstone modes propagating only inside the vortex;  outside the vortex they become massive and do not propagate. 
However,  it is by now well understood that these vortices Abelianize dynamically. The fluctuations of their orientational modes  become strong at long distances (low energies) and they lose effectively their orientational moduli.  At low energies they are at one of the $N$ ``vortex vacua'',  and the kinks connecting different vortices are Abelian monopoles.  

As we know that in the $4D$  gauge theories there appear light monopoles which carry genuine non-Abelian charges,
there must be semi-classical vortices which carry non-Abelian orientational moduli, which do not dynamically Abelianize.  
Such a model has been constructed recently \cite{DKO}.
The underlying theory is the same softly broken ${\cal N}=2$  SQCD,
Eq.~(\ref{lagrangianGeneral}),  Eq.~(\ref{lagquark}),  
but this time  we tune the bare quark masses as 
\[  m_{1}= \ldots = m_{n} = m^{(1)}; \quad m_{n+1}= m_{n+2}= \ldots = m_{n+r}= m^{(2)}\, , \qquad N=n+r\;;  
 \] 
\be    n\, m^{(1)} +  r\, m^{(2)} =0\;, \label{masses} \ee
or
\[   m^{(1)}=  \frac{r\, m_{0}}{\sqrt{ r^{2} + n^{2}}}, \quad   m^{(2)}= - \frac{ n \, m_{0}}{\sqrt{ r^{2} + n^{2} }},
\label{masscond}\]
and their magnitude is taken as 
 \be   |m_{0}| \gg  |\mu|  \gg  \Lambda\,.\label{masssemi}
\ee
  The adjoint scalar VEV can be taken to be 
\be \brc \Phi \ckt   =  - \frac{1}{\sqrt{2}}  \left(\begin{array}{cc}m^{(1)} \, {\mathbbm 1}_{n \times n} & {\bf 0} \\ {\bf 0}  & m^{(2)} \, {\mathbbm 1}_{r \times r}  \end{array}\right)  \label{ourvac}
\ee
Below the mass scale  $v_{1} \sim   |m_{i}|$   the system thus  reduces to a gauge theory with gauge group 
\be   G =   \frac {SU(n) \times SU(r) \times U(1)}{{\mathbbm Z}_{K}}\,,   \quad K = {\rm LCM} \, \{n,r\}\;    \label{Gaugegr} 
\ee
 where $K$ is the least common multiple of $n$ and $r$.   The higher $n$ color components of the first $n$ flavors  (with the bare mass $m^{(1)}$)   remain massless, as well as the lower $r$ color components of the last $r$ flavors
 (with the bare mass $m^{(2)}$):  they will be denoted as $q^{(1)}$ and $q^{(2)},$
 respectively. 
 
 Our low energy system  then is: 
\bqa  && {\cal L} =  - \frac{ 1 }{4 g_{0}^2}  F_{\mu \nu}^{0\, 2}  - \frac{ 1 }{4 g_{n}^2}  F_{\mu \nu}^{n\, 2} - \frac{ 1 }{4 g_{r}^2}  F_{\mu \nu}^{r\, 2}   +    \frac { 1}{ g_{0}^2}  |{\cal D}_{\mu} \Phi^{(0)}|^2 +  \frac { 1}{ g_{n}^2}  |{\cal D}_{\mu} \Phi^{(n)}|^2 +   \non \\
&& + \frac { 1}{ g_{r}^2}  |{\cal D}_{\mu} \Phi^{(r)}|^2 +     \left|{\cal D}_{\mu}  q^{(1)} \right|^2 + \left|{\cal D}_{\mu} \bar{\tilde{q}}^{(1)}\right|^2  + 
  \left|{\cal D}_{\mu}  q^{(2)} \right|^2 + \left|{\cal D}_{\mu} \bar{\tilde{q}}^{(2)}\right|^2     -    V_D -  V_F, 
\non  \eea
plus fermionic terms,  
where  $V_{D}$ and $V_{F}$ are the $D$-term and $F$-term potentials.    The $D-$term potential $V_{D}$ has the form, 
 \[ V_{D}   =          \frac { 1}{ 8 }  \, \sum_A  \left(\Tr\, t^A \,[
 \,  \frac { 2}{ g^2 }  \, [\Phi,   \Phi^\dagger]  +  
\sum_{i}(Q_i   Q_i^\dagger -  {\tilde Q}_i^\dagger   {\tilde Q}_i )\,] \right)^2;
\]
where  the generators $A$ takes the values  $0$ for $U(1)$,  $a=1,2, \ldots, n^{2}-1$ for $SU(n)$ and  $b=1,2, \ldots r^{2}-1$ for $SU(r)$.
The mass matrix takes the form 
\[    M=  \left(\begin{array}{cc}m^{(1)} \, {\mathbbm 1}_{n \times n} & {\bf 0} \\ {\bf 0}  & m^{(2)} \, {\mathbbm 1}_{r \times r}  \end{array}\right)
\]
and  the (massless) squark fields are 
\[   Q(x) =   \left(\begin{array}{c c}  q^{(1)}(x) _{n\times n}  & 0 \\ 0  &  q^{(2)}(x)_{r \times r}\end{array}\right)\,,\quad   {\tilde Q}(x) =   \left(\begin{array}{c c}  {\tilde q}^{(1)}(x) _{n\times n}  & 0 \\ 0  &  {\tilde q}^{(2)}(x)_{r \times r}\end{array}\right)\,, \]
if written in a color-flavor mixed matrix notation.  


The VEVs of the adjoint scalars are  given by 
\be   \brc \Phi^{(0)} \ckt =-  m_{0}, \qquad \brc \Phi^{(a)} \ckt = \brc \Phi^{(b)} \ckt   =0,\; \label{adjVev}
\ee
while the squark VEVs are given     by
\[    
\brc Q \ckt =   \left(\begin{array}{c c}  v^{(1)} \, {\mathbbm 1}_{n\times n}  & 0 \\ 0  &  - v^{(2)\, *} \, {\mathbbm 1}_{r\times r}  \end{array}\right)\,,\quad \brc {\tilde Q} \ckt =   \left(\begin{array}{c c}  v^{(1)\, *} \, {\mathbbm 1}_{n\times n}  & 0 \\ 0  &     v^{(2)} \, {\mathbbm 1}_{r\times r}  \end{array}\right)\,,
\label{QkVEV}\]
with
\[ 
 | v^{(1)}|^{2} +    | v^{(2)}|^{2}  = \sqrt{\frac {n+r}{n\, r}} \,  \mu \, m_{0}\,  \;. 
\]
There is a continuous vacuum degeneracy;  we assume that 
\[    v^{(1)} \ne 0; \qquad  v^{(2)} \ne 0\;,
\]
in the following. 

Non-Abelian vortices exist in this theory as the vacuum breaks the gauge group $G$ (Eq.~(\ref{Gaugegr})) completely, leaving at the same time a color-flavor diagonal symmetry 
\be   [ SU(n) \times SU(r) \times U(1) ]_{C+F}  
\label{colorflavor}\ee
 unbroken.   The full global symmetry, including the overall global $U(1)$ is given by 
 \be   U(n) \times U(r)\;.  \label{notethat}
 \ee
   The minimal vortex in this system corresponds to the smallest nontrivial loop in the $G$ group space,  Eq.~(\ref{Gaugegr}).
It is the path in the $U(1)$ space
\[   \left(\begin{array}{cc} e^{i \alpha r}  {\mathbbm 1}_{n \times n} & 0 \\0 & e^{i \alpha n} {\mathbbm 1}_{r \times r }\end{array}\right)\,,\quad \alpha  : 0 \to    \frac{2\pi }{ n\, r },
\]
that is, 
\[  {\mathbbm 1}_{N \times N} \to    {\mathbbm Y}, 
\qquad      {\mathbbm Y} = \left(\begin{array}{cc}e^{2 \pi i/n } {\mathbbm 1}_{n \times n} & 0 \\0 & e^{ 2 \pi i / r} {\mathbbm 1}_{r \times r }\end{array}\right)\,, 
\]
followed by a path 
 in the $SU(n) \times SU(r)$ manifold 
 \[   {\mathbbm 1}_{n\times n} \to  {\mathbbm Z}_{n}=  e^{
-\frac{2 \pi i}{n} } {\mathbbm 1}_{n\times n} ; \qquad 
{\mathbbm 1}_{r\times r} \to   {\mathbbm Z}_{r}= e^{-\frac{2 \pi i}{r} } {\mathbbm 1}_{r\times r} ;\]
 back to the unit element. For instance one may choose 
  ($\beta:0 \to 2\pi; \gamma:0\to 2\pi$)
 \[  \left(\begin{array}{cc}e^{i\beta (n-1) /n} & 0 \\ 0 & e^{-i \beta /n}\, {\mathbbm 1}_{(n-1)\times (n-1)}\end{array}\right)\;;\quad 
 \left(\begin{array}{cc}e^{i\gamma (r-1) /r} & 0 \\ 0 & e^{-i \gamma /r}\, {\mathbbm 1}_{(r-1)\times (r-1)}\end{array}\right)\;.
  \]
   As   
 \[  {\mathbbm Y}^{K}  =   {\mathbbm 1}_{N \times N},  \quad K = {\rm LCM} \, \{n,r\}\;
 \]
 it follows that the tension (and the winding) with respect to the $U(1)$ is   $\frac{1}{K} $ of that in the standard ANO  vortex.  

\begin{figure}
\begin{center}
\includegraphics[width=2.5 in]{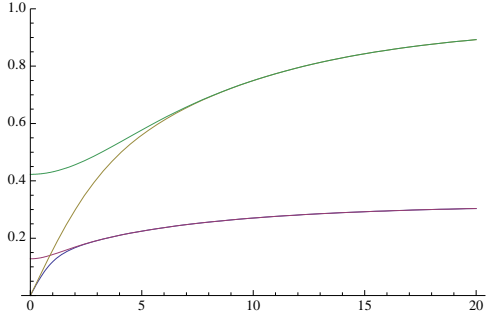}
\caption{\footnotesize Numerical result for the profile functions $f_{1,2}, g_{1,2}$ as functions of the radius $\rho$, for $SU(3)\times SU(2)\times U(1)$  theory.  The coupling constants and the ratio of the VEVs are taken to be $g_{0}=0.1$, $g_{3}=10$, $g_{2}=1$, $|v_{2}|/ |v_{1}|=3$.  }
\label{Vortexprofile}
\end{center}
\end{figure}

The squark fields trace such a path asymptotically, i.e., far  from the vortex core,
as one goes around the vortex;    at finite radius the vortex has, for instance, the form,  
{\small
\[   q^{(1)}  =   \left(\begin{array}{c c    }  e^{i \phi} \,f_{1}(\rho)  &  0   \\
0 &   f_{2}(\rho)\, {\mathbbm 1}_{(n-1)\times (n-1)}  
 \end{array}\right), \quad 
 {\tilde q}^{(2)}  =   \left(\begin{array}{c c    }  e^{i \phi} \,g_{1}(\rho)  &  0   \\
0 &  g_{2}(\rho)\, {\mathbbm 1}_{(r-1)\times (r-1)}  
 \end{array}\right), 
\label{vortexconf} \]
}
 where $\rho$ and $\phi$ stand for the polar coordinates in the plane perpendicular to the vortex axis,  $f_{1,2}, g_{1,2}$ are profile functions.  The adjoint scalar fields $\Phi$ are taken to be equal to their VEVs, Eq.~(\ref{adjVev}).  They are accompanied by the appropriate gauge fields so that the  tension is finite.   The BPS equations for the squark and gauge fields, and the properties of their solutions are discussed in the original paper \cite{DKO}.  The behavior of numerically integrated vortex profile functions $f_{1,2}, g_{1,2}$ is illustrated in Fig.~\ref{Vortexprofile}. 

   We note here only that the necessary boundary conditions on the squark profile functions have the form, 
   \[     f_{1}(\infty)=  f_{2}(\infty) = v^{(1)}, \qquad  g_{1}(\infty)= g_{2}(\infty) = v^{(2)}, 
   \]
   while at the vortex core, 
   \[  f_{1}(0)=0, \quad  g_{1}(0) =0, \qquad   f_{2}(0) \ne 0, \quad  g_{2}(0) \ne 0,
  \label{fundamental} \]
   The most important fact about these minimum vortices  is that  one of the $q^{(1)}$ {\it and}  one of the ${\tilde q}^{(2)}$ fields must necessarily wind at infinity, simultaneously.  
   As the individual vortex breaks the (global) symmetry of the vacuum as 
\be  [ SU(n) \times SU(r) \times U(1) ]_{C+F}  \to  SU(n-1)  \times SU(r-1)  \times U(1)^{3}, 
\label{smaller}  \ee
the vortex acquires  Nambu-Goldstone modes parametrizing
\be   CP^{n-1} \times CP^{r-1}\;:         
\ee
they transform under the exact color-flavor symmetry $SU(n) \times SU(r)$  as the 
bi-fundamental  representation, $({\underline n}, {\underline r})$.
Allowing the vortex orientation to fluctuate along the vortex length and in time, we get a $CP^{n-1} \times CP^{r-1}$  two-dimensional sigma model as an effective Lagrangian describing them.  The details have been worked out in \cite{SY,HT2}  and  need not be repeated here.

\begin{figure}
\begin{center}
\includegraphics[width=3in]{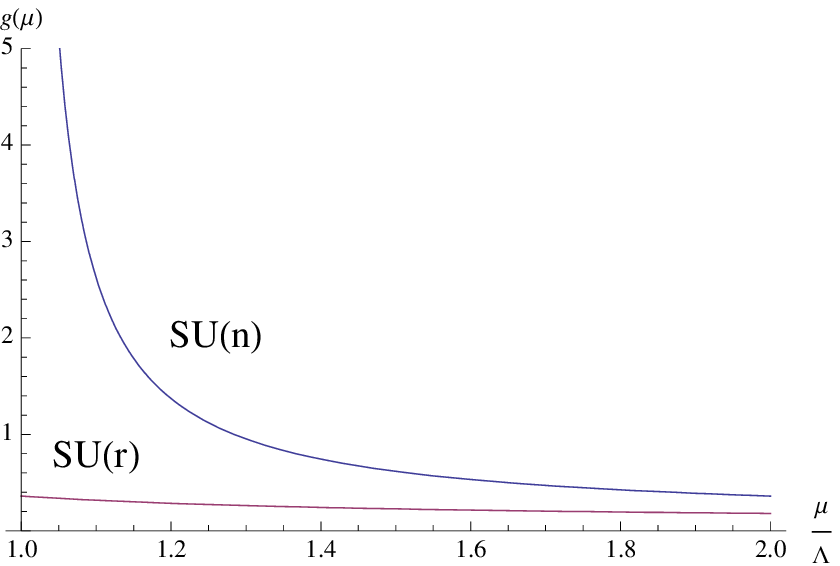}
\caption{ }
\label{sunsur}
\end{center}
\end{figure}

  Let us assume without losing generality that  $n > r$, excluding the special  case of $r=n$  for the moment.  As has been shown in \cite{SY,HT2} the coupling constant of the $CP^{n-1}$  sigma models grows precisely as the  coupling constant of the $4D$ $SU(n)$ gauge theory.   At the point the $CP^{n-1}$  vortex moduli fluctuations become strong and 
the dynamical scale $\Lambda$ gets generated, with vortex kinks (Abelian monopoles) acquiring mass of the order of  $\Lambda$, 
the vortex still carries the unbroken $SU(r)$  fluctuation modes ($CP^{r-1}$),  as 
the $SU(r)$ interactions are still weak.  See Fig.~\ref{sunsur}.
  Such a vortex will carry one of the $U(1)$ flux 
arising from the dynamical breaking of $SU(n) \times U(1) \to  U(1)^{n}$, as well as an $SU(r)$ flux.  
As these vortices end at a massive monopole (arising from the high-energy gauge symmetry breaking, Eq.~(\ref{ourvac})),  the latter necessarily carries a non-Abelian continuous moduli, whose points transform as in
the fundamental representation of 
$SU(r)$.  This can be  interpreted as the  (electric description of)  dual gauge $SU(r)$ system observed in the infrared limit of the $4D$ SQCD \cite{APS,CKM}.  

The special case $r=1$ corresponds to the $U(N)$ model  \cite{ABEKY,SY,HT2,Eto:2006pg}, discussed earlier, where 
vortices dynamically Abelianize. Note that in that case 
the corresponding semiclassical vacuum is the one labeled by $r=N$. These facts are perfectly consistent with the classical-quantum vacuum matching conditions \cite{BKM}.  
 Eq.~(\ref{notethat}) also perfectly matches the full quantum result \cite{CKM},  for generic $r$.

  \section{Birth of the dual gauge group}
 
Thus  vortices having non-Abelian moduli, which do not dynamically completely Abelianize, can be constructed.   Semi-classically, they are simply vortices carrying the $SU(n) \times SU(r) \times U(1)$ color-flavor flux.  More precisely,  they carry the Nambu-Goldstone modes 
\[  CP^{n-1}\times CP^{r-1}\;,
\]
resulting from the partial breaking of the $SU(n) \times SU(r) \times U(1)$  global symmetry to $SU(n-1) \times SU(r-1) \times U(1)^{3}$ by the vortex.    For  $n>r$,   $CP^{n-1}$ field fluctuations propagating along the vortex length  become strongly coupled in the infrared, the $SU(n) \times U(1)$ part dynamically Abelianizes;  the vortex however still carries weakly-fluctuating $SU(r)$ flux modulations.  In our  theory  where  $SU(n) \times SU(r) \times U(1)$ model arises as the low-energy approximation of an underlying $SU(N)$ theory,  such a vortex is not stable.  When the vortex ends at a monopole,  its  $CP^{r-1}$ orientational modes  turn  into the dual $SU(r)$ color modulations of the monopole.

\section*{Acknowledgements}

I am grateful to the organizers of the Conference  "30 Years of Mathematical Methods in High Energy Physics -- in honor of Prof. T. Eguchi's 60th Birthday" (Kyoto, March 2008), especially Hirosi Ooguri, for inviting me to participate and contribute in celebrating Tohru Eguchi's 60th birthday, which was a great pleasure for me and was a very stimulating event for everybody present.  The first part of the talk presented here is a state-of-the-art summary of the work done in collaboration with many friends,  R. Auzzi, S. Bolognesi, D. Dorigoni,  J. Evslin, M. Eto, L. Ferretti, S.B. Gudnason, T. Fujimori, M. Nitta, K. Ohashi, W. Vinci, N. Yokoi, among them.

%

\end{document}